# Unraveling a Structure-Property Relationship for Methylammonium Lead/Tin Trihalide Organic-Inorganic Hybrid Perovskite Solar Cells

Arpita Varadwaj,[a,b] Pradeep R. Varadwaj,[a,b,*] Koichi Yamashita[a,b]

[a]Department of Chemical System Engineering, School of Engineering, The University of Tokyo 7-3-1, Hongo, Bunkyo-ku, Japan 113-8656
[b]CREST-JST, 7 Gobancho, Chiyoda-ku, Tokyo, Japan 102-0076

Abstract

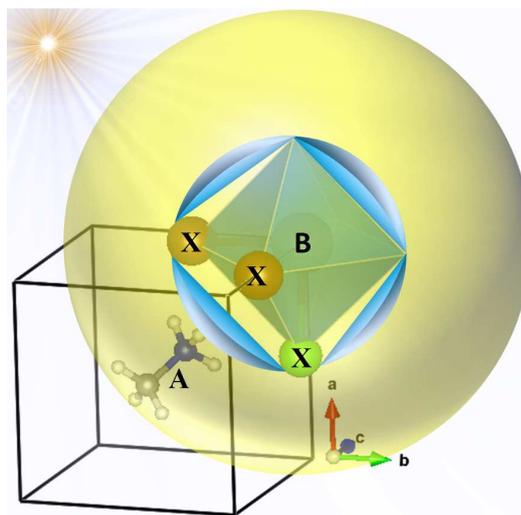

The esoteric importance of intermolecular hydrogen bonding interaction to design novel methylammonium trihalide organic-inorganic hybrid perovskite solar cell materials is uncovered, establishing the unified structure-property relationship between the calculated Y•••H intermolecular hydrogen bonding distance and the experimentally reported onset of optical absorption (bandgap) for the bulk geometries of the ten-membered methylammonium lead trihalide ($CH_3NH_3PbY_3$) perovskite solar cell series, where Y = X (X = Cl, Br, I) and the mixed halogen derivatives. The same relationship is also revealed for the ten-membered methylammonium tin trihalide ($CH_3NH_3SnY_3$) perovskite solar cell series. The relationship unequivocally demonstrates that the intermolecular hydrogen bonding interaction does not only enforce the aforesaid materials to become functional for optoelectronic application, but also serve as an asset to partially address the often debated question what is the role played by the $CH_3NH_3^+$ organic cation.

[*] Corresponding Author's E-mail Addresses: pradeep@t.okayama-u.ac.jp (PRV);
varadwaj.arpita@gmail.com (AV); yamasita@chemsys.t.u-tokyo.ac.jp (KY)

1. **Introduction**

Methylammonium Lead (or Tin) Trihalide ($CH_3NH_3PbX_3$, $CH_3NH_3SnX_3$; X = Cl, Br, I) organic-inorganic hybrid perovskite thin films are advanced high-ranking semiconductors discovered till date.[1] These are debated as suitable nanomaterials for designing high performance photovoltaic devices. This is so because these are having unbelievably large potential to convert the Sun's photon energy into electricity,[2] an ultimate source for future energy from renewables. Efforts to examine this specific light conversion feature by diverse research groups have succeeded in leaps and bounds, reaching to a National Renewable Energy Laboratory certified Power Conversion Efficiency (PCE) of 22.1%.[3] Demonstrations to further boost photovoltaic device PCE up to 35% are also documented.[4]

The nature of the halogen content, and the nature of the organic and inorganic cations jointly play the crucial role in generating the novel opto-electronics and -physics of the trihalide based perovskite solar cell materials.[5] These, together with the nature of the oxidation state of the participating species, the cation size, the electronegativity, the Goldschmidt tolerance factor, and the octahedral factor, among others, are the determinants responsible for explaining the formability and octahedral stability of the trihalide perovskite solar cells, as well as in generating their novel properties.[5,6]

It is known that the strength of bandgap $E_g$ depends on the halide orbital energy, in which, both of these are related to the size of the halide atom. Lattice parameters are correlated with the size of the halide ion for trihalide based perovskites. Hence a linear dependence between lattice parameter and bandgap is generally expected since bandgap energy is inversely proportional to the square of lattice constant.[7] This view is perhaps in accord with Castelli *et al.*,[8] who have demonstrated in a recent study that it does not matter what is the temperature phase of the system (orthorhombic, tetragonal or cubic) the decrease of the bandgap for a trihalide-based perovskite series with a given set of A- and B-site cations is a consequence of the increase in the size of the halogen in the halogen series from Cl to Br to I (or, the decrease of the electronegativity of the corresponding species in this order: Cl > Br > I) that assists in the decrease in the lattice constant. However, the authors aware that the structural stability of the trihalide perovskite solar cell is achieved due to the marriage between the inorganic and organic species, with the 3D supramolecular geometry (e.g., thin films) is simply the periodic analogue of the bulk. The marriage is strong in the orthorhombic phase, and medium-to-weak in the tetragonal and pseudocubic phases.[9] To this end, it is genuine to ask the most fundamental question whether or not there can be any such dependence expected between the Y•••H hydrogen bond distance and bandgap for each of the two perovskite series, $CH_3NH_3PbY_3$ and $CH_3NH_3SnY_3$, where Y = $I_{(3-x)}Br_{x=1-3}$, $I_{(3-x)}Cl_{x=1-3}$, $Br_{(3-x)}Cl_{x=1-3}$, and IBrCl.

## 2. Computational details

The study has undertaken periodic density functional theory calculation to examine the structural aspects of the relaxed bulk geometries of the ten-membered methylammonium lead trihalide $CH_3NH_3PbY_3$ perovskite solar cell series. Similar calculations were performed to examine the same property for the other ten members of the methylammonium tin trihalide $CH_3NH_3SnY_3$ perovskite solar cell series. The Perdew-Burke-Ernzerhof (PBE) exchange-correlation functional, together with the projector augmented wave (PAW) potentials for all atoms, the cut-off energy for the plane wave pseudopotential of 500 eV, a 6×6×6 Γ-centered k-point sampling of the Brillouin zone, and the tetrahedron method with Blöchl corrections for the Brillouin zone integrations, as implemented in VASP, [10] were set for the relaxations of the geometries of these two series in the pseudocubic phase. Convergence criteria for energy and force were set to $1 \times 10^{-5}$ eV and 0.01 eV Å$^{-1}$, respectively.

Unless otherwise stated, MA will be referred hereafter as either $CH_3NH_3^+$, or MA.

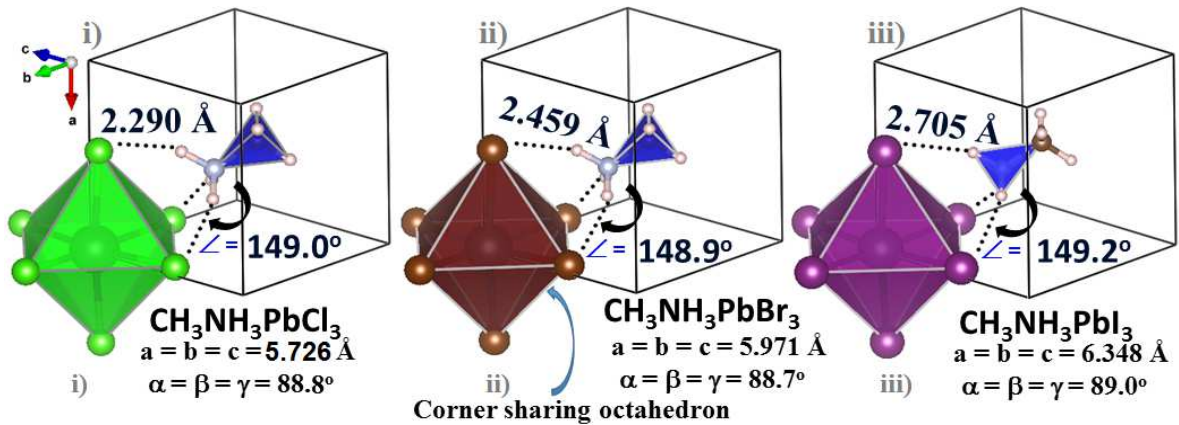

Fig. 1: Comparison between some selected geometrical properties for the i) $CH_3NH_3PbCl_3$, ii) $CH_3NH_3PbBr_3$ and iii) $CH_3NH_3PbI_3$ bulks. The lattice constants for each system are given.

## 3. Results and Discussion

While the geometries of the twenty perovskite solar cell bulks (10 Pb-based and 10 Sn-based) are relaxed using periodic DFT by varying the halogen in all possible combinations, Fig. 1 illustrates only three of them, viz. $CH_3NH_3PbX_3$ (X = Cl, Br, I), as examples. The calculated lattice constants shown are in good agreement with those reported experimentally, a = 6.315(3) for

$CH_3NH_3PbI_3$, [9a)] a = 5.93129(4) for $CH_3NH_3PbBr_3$, [11] and 5.68415(6) Å for $CH_3NH_3PbCl_3$. [11] In all the cases, the copulation between the $PbY_3^-$ and $CH_3NH_3^+$ fragments, as well as that between $SnY_3^-$ and $CH_3NH_3^+$, is found to be highly attractive, which are evolved in the $CH_3NH_3PbY_3$ and $CH_3NH_3SnY_3$ systems as the Y•••H–N geometrical motifs, where the three dots represent the intermolecular hydrogen bonding interaction.

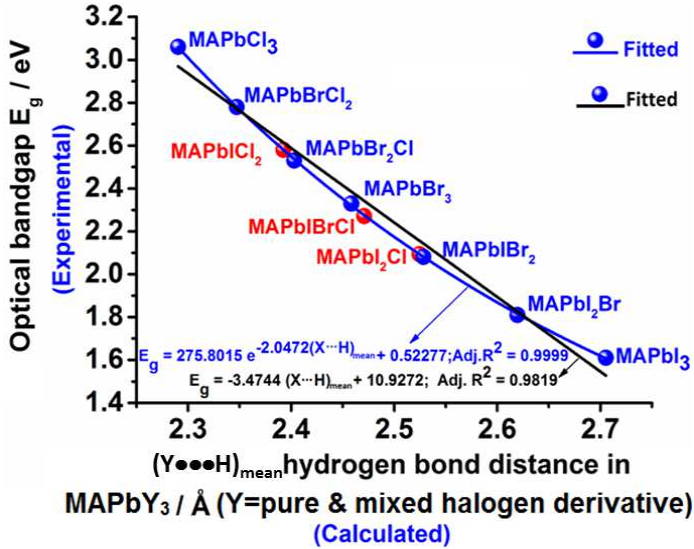

Fig. 2: Dependence of experimentally reported optical bandgap $E_g$ on the calculated intermolecular hydrogen bonding interaction distance $(Y•••H)_{mean}$ for the ten-membered $MAPbY_3$ series (Y = pure as well as mixed halogen derivatives, MA = $CH_3NH_3^+$). The green and red spheres represent experimental and predicted $E_g$, respectively. The blue and red spheres represent experimental and predicted $E_g$, respectively.

Fig. 2 illustrates the relationship between the experimentally available $E_g$ and the calculated intermolecular hydrogen bonding distance for the $CH_3NH_3PbY_3$ perovskite solar cell bulks. The experimental $E_g$ values represent the blue spheres, which involve $MAPbI_3$ (1.61 eV), $MAPbBr_3$ (2.33 eV), $MAPbCl_3$ (3.06 eV), $MAPbI_2Br$ (1.81 eV), $MAPbIBr_2$ (2.08 eV), $MAPbBr_2Cl$ (2.53 eV), and $MAPbBrCl_2$ (2.78 eV). [12] These data were separately fitted to two different equations, which were $y = a + b{\times}x$, and $y = a_0 + b_0{\times}e^{-c_0 x}$, where $a$ and $b$ are the constants of the former equation, and $a_0$, $b_0$ and $c_0$ are the constants of the latter equation. The regression analysis is presented in Fig. 2. It is evident from the graph that fitting of the former and latter equations have produced regression coefficients, which are as good as 0.982 (black line) and 0.999 (blue curve), respectively, meaning that the data fitted to the curve is better with the latter than that with the former equation. We thereby have used the fitting parameters of the latter equation ($y = a_0 + b_0{\times}e^{-c_0 x}$) to predict the bandgaps for the remaining three missing members of the $CH_3NH_3PbY_3$ family. These include $MAPbICl_2$ (2.58 eV), $MAPbIBrCl$ (2.27 eV) and $MAPbI_2Cl$ (2.09 eV), in which, the $E_g$ for these three are yet to be experimentally examined.

Similarly, as a second example, and to show the reliability of the above relationship between the bandgap and hydrogen bond distance we extended our study to perform the same analysis for the $CH_3NH_3SnY_3$ perovskite solar cell series. Fig. 3 presents the desired graph, illustrating the dependency of the experimentally available $E_g$ on the calculated $(Y\bullet\bullet\bullet H)_{mean}$ for the $CH_3NH_3SnY_3$ perovskite solar cell series. But, in this case, the $E_g$ values were experimentally available only for five systems. These include the $MASnI_3$ (1.30 eV), $MASnI_2Br$ (1.56 eV), $MASnBr_3$ (2.15 eV), $MASnIBr_2$ (1.75 eV), and $MASnCl_3$ (2.80 eV). [13] Because the fitting of the Eqn $y = a_0 + b_0 \times e^{-c_0 x}$ to the $CH_3NH_3PbY_3$ data set has produced better regression coefficient compared to the other (see discussions above), we have used the experimentally available $E_g$ and currently computed $(Y\bullet\bullet\bullet H)_{mean}$ data for the $CH_3NH_3SnY_3$ systems to fit them to the former Eqn. As shown from by Castelli *et al.*,[8] as the difference between the $E_g$ values for the MASnIBr$_2$ and MASnI$_2$Cl systems are seemingly not very marked, with the former marginally smaller than the latter.

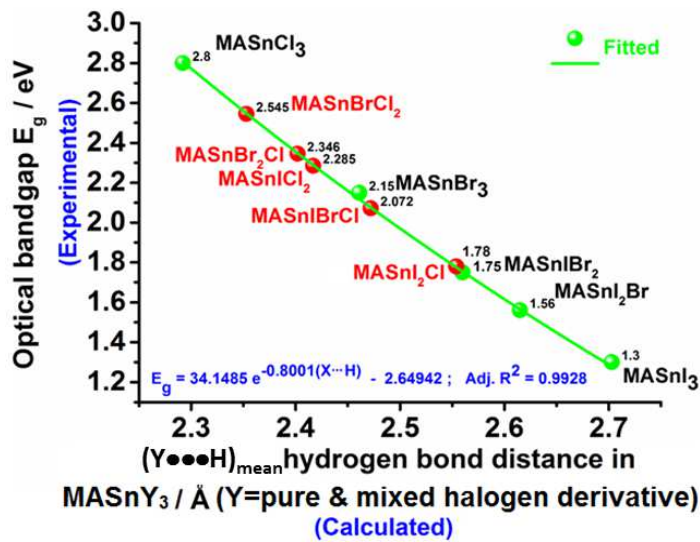

Fig. 3: Dependence of experimentally reported optical bandgap $E_g$ on the calculated intermolecular hydrogen bonding interaction distance $(Y\bullet\bullet\bullet H)_{mean}$ for the ten-membered $MASnY_3$ series (Y = pure as well as mixed halogen derivatives, MA = $CH_3NH_3^+$). The green and red spheres represent experimental and predicted $E_g$, respectively. Exact $E_g$ values in eV (both experimental and predicted) are depicted for clarification.

Fig. 3, the fitting procedure has resulted in Adj. $R^2 \approx 0.993$, which is indeed comparable with that found for the $CH_3NH_3PbY_3$ set (cf. Fig. 2). However, the constants resulted from the current fit are not same as those found for the $CH_3NH_3PbY_3$ set, even though the same equation was used for fitting the data points. This may not be very unsurprising because the metal ion constituent for each perovskite series was different, so are the bandgaps and intermolecular hydrogen bonding distances. Nevertheless, based on the constants obtained from the fitting procedure (cf. Fig. 3), the bandgaps for the remaining members of the $CH_3NH_3SnY_3$ family are predicted. These include the $MASnI_2Cl$, $MASnClBrI$, $MASnCl_2I$, $MASnClBr_2$, and $MASnCl_2Br$ that have bandgaps of 1.78, 2.07, 2.28 and 2.35 eV, respectively, which are visible in Fig. 3 as tiny spheres in red. The order of preference in $E_g$ for all the ten members of the family is not strictly following the one proposed

In summary, we have performed density functional theory calculation employing periodic boundary condition to determine the intermolecular hydrogen bonding geometries for two different trihalide based perovskite solar cell series, one comprising the CH$_3$NH$_3$PbY$_3$, and the other the CH$_3$NH$_3$SnY$_3$. The intermolecular hydrogen bonding distances obtained from this calculation for all the twenty members of the lead(II) and tin(II) families and the experimental bandgap data reported for most of the members of both the families were used in 2D curve fitting separately to obtain the desired parameters of the Eqn: $y = a_0 + b_0 \times e^{-c_0 x}$. While the nature of the fitting parameters for each series was different, these were found very useful in predicting separately the bandgaps of the remaining missing members of each of the lead and tin trihalide perovskite solar cells families yet to be experimentally examined. This result enabled us to understand that the novel optoelectronic properties for which the high-performance trihalide perovskite soft nanomaterials are recognized today to be ideal for device applications are certainly the consequence of intermolecular hydrogen bonding interactions, which are attractive, directional, and plays the role of a soft gel-like adhesive, with the latter essentially important for structural manipulations and supramolecular designs of novel functional materials. This perspective is indeed reflected in the unified structure-property relationship established in this work between the intermolecular distance of separation and the optical bandgap for each perovskite solar cell series, CH$_3$NH$_3$PbY$_3$ or CH$_3$NH$_3$SnY$_3$. The relationship suggests that with the knowledge of the hydrogen bonding parameter, such as the intermolecular distance of separation, it is possible even to infer and evaluate the bandgap of any hybrid organic-inorganic trihalide perovskite solar cell in a series of given type, similarly as how it is achieved for a few missing members of the two perovskite solar cell series investigated. The study enlightened the fact that the fundamental knowledge of intermolecular hydrogen bonding is potentially and essentially pragmatic in the design of new novel materials of similar kind and beyond. Also, this paper answered to some extent the question what is the role of the organic cation in the CH$_3$NH$_3$PbY$_3$ and CH$_3$NH$_3$SnY$_3$ perovskite solar cells. The discussions provided within the context of this commutation will undoubtedly be serendipitous for the future development of advanced trihalide based perovskite nanomaterials for energy applications and beyond.